\documentclass[12pt]{amsart}

\usepackage{graphicx}

\begin{document}

\title {Hysteresis effects in Bose-Einstein condensates}

\author {Andrea Sacchetti}

\address {Faculty of Sciences - 
University of Modena e Reggio Emilia - 
Via Campi 213/B, I--41100 Modena, Italy}

\date {\today}

\email {Andrea.Sacchetti@unimore.it}

\begin {abstract}
Here, we consider damped two-components Bose-Einstein condensates with many-body interactions. \ We show that, when the external trapping potential has a double-well shape and when the nonlinear coupling factors are modulated in time, hysteresis effects may appear under some circumstances. \ Such hysteresis phenomena are a result of the joint contribution between the appearance of saddle node bifurcations and damping effect. 

PACS number(s): {05.45.-a, 02.30.Oz, 03.65.Ge, 03.75.Lm}
\end{abstract}

\maketitle

Atomic Bose-Einstein condensates (BECs) at zero temperature are described by means of nonlinear Schr\"odinger equations of the type
\begin{eqnarray}
i \hbar \frac {\partial \psi}{\partial t} = H_0 \psi + g_2 |\psi |^{2 } \psi + g_3 |\psi |^{4 } \psi + \ldots \label {equation1}
\end{eqnarray}
where $H_0$ represents the Hamiltonian of a single trapped atom and the nonlinear term $ |\psi|^{2r} $, $r=1,2,\ldots $, is the $(r+1)$-body contact potential with coupling factor $g_{r+1}$ \cite {Kohler}. \ In fact, BECs strongly depend by interatomic forces and the binary coupling term $g_2 |\psi|^2 \psi$ usually represents the dominant nonlinear term; when the higher nonlinear terms are neglected then equation (\ref {equation1}) takes the form of the well-known Gross-Pitaevskii equation \cite {PitStr}. \ The coupling factor of the binary nonlinear term is given by $g_2 = {\mathcal N} 2 \pi \hbar^2 a /m$ where $m$ is the mass of the atoms, ${\mathcal N}$ is the total number of particles of the condensate, and $a$ is the scattering length; for higher nonlinearity terms some expressions of the coupling factor have been recently proposed \cite {Braaten}. \ Recent experiments \cite {Cornish} have shown that the scattering length $a$ can be changed, and, in fact, the $2$-body coupling factor $g_2$ can be tuned to be zero in the case of polar molecules in optical lattices driven by microwave fields \cite {Buchler}. \ In such a case, the $3$-body and, in general, $(r+1)$-body interaction becomes significant \cite {Roberts, Gammal} and thus the Gross-Pitaevskii equation becomes inadequate in order to describe BECs.

The basic properties of BECs with many-nody interactions described by equation (\ref {equation1}), where many nonlinear terms are simultaneously considered, are far to be well understood. \ In order to understand some fundamental features of BECs with many-body interactions in this paper we follow the approach proposed by \cite {Fersino}, that is we are interested to the $(r+1)$-body intereaction in its own. \ Therefore, we restrict ourselves to the basic nonlinear Schr\"odinger equation representing the properties of exactly $(r+1)$-body contact interaction of BECs at zero temperature
\begin{eqnarray}
i \hbar \frac {\partial \psi}{\partial t} = H_0 \psi + g_{r+1} |\psi |^{2 r } \psi , \  \ \| \psi \| =1 \, , \label {equation2}
\end{eqnarray}
which depends on the Hamiltonian 
\begin {eqnarray*}
H_0 = - \frac {\hbar^2}{2m} \sum_{j=1}^n \frac {\partial^2}{\partial x_j^2} + V(x)
\end {eqnarray*}
of a single atom in dimension $n$ with trapping potential $V(x)$, as well as on the $(r+1)$-body coupling factor $g_{r+1}$. 

It is worth mentioning also the fact that equation (\ref {equation2}) with nonlinearity corresponding to the power-law $|\psi |^{2r}$, where the parameter $r$ takes any positive real value,  is used in other contexts, including semiconductors \cite {Mihalache} and nonlinear optics \cite {Zakharov, Snyder, Christian}. \ Furthermore, even if in most of the applications the parameter $r$ takes only integer and positive values, here we take that $r$ can assume non integer values too, as considered in \cite {Smerzi}.

Since the wavefunction $\psi$ is assumed to be normalized to $1$ then the coupling factor $g_{r+1}$ depends, in addition to the physical parameters of the problem such as the scattering length and the mass of the particles of the condensate, on the total number ${\mathcal N}$ of the particles of the condensate. 

Recent experiments have shown that the total number ${\mathcal N}$ of particles that participate to the condensate can be adiabatically modulated in time by means of a suitable  time-dependent combination of optical and magnetic forces \cite {Stamper,Ma}. \ Thus, we can consider the case where the nonlinear coupling factor in equation (\ref {equation2}) is a given function which slowly depends on the time $t$ 
\begin {eqnarray*}
g_{r+1} := g_{r+1} (t) \, . 
\end {eqnarray*}
Another way to produce a time-dependent coupling factor consists of tuning the scattering length \cite {KK1}. \ In a previous theoretical papers by Pelinovsky, Kevrekidis and Frantzeskakis \cite {PKF} the Gross-Pitaevskii equation with a periodically varying nonlinearity coupling factor has been considered and it has been shown a good agreement between solutions of the averaged and full equations. 

In this paper we show that the modulation of the nonlinearity coupling factor may give rise to hysteresis phenomenon for two-components BECs, where the the external trapping potential $V(x)$ has a double-well shape \cite {Salasnich}. \ In fact, hysteresis effects have been already seen in rotating BECs, in particular the number of vortices appearing in a rotate BEC depends on the rotation history of the trap, in addition to the number of vortices initially present in the condensate \cite {Jackson}; see also the theoretical analysis in \cite {Rozanov}.

In particular, here we show that, in the semiclassical limit of $\hbar$ small enough, for BEC's equation (\ref {equation2}) with power law $|\psi |^{2r}$ in a double well trapping potential and with a slow modulation, with respect to the beating period between the two wells, of the nonlinear  coupling factor $g_{r+1}(t)$ then an hysteresis effect appears provided that $r$ is bigger than the critical value 
\begin{eqnarray}
r_{threshold} = (3+\sqrt {13})/{2} \, . \label {equation3}
\end{eqnarray}
It is worth mentioning the fact that this result holds true for both attractive (e.g. $g_{r+1}<0$) and repulsive (e.g. $g_{r+1}>0$) nonlinearities; however, just for argument's sake, we restrict ourselves to the attractive case \cite {attractive}.

The hysteresis effect is strictly close to the appearance of spontaneous symmetry breaking phenomenon (SSBP) related to saddle point nodes. \ In fact, for BECs with $(r+1)$-body interaction governed by equation (\ref {equation2}) has been recently seen \cite {Sacchetti} that SSBT appears when the nonlinearity power $r$ is bigger than  $r_{threshold}$.  \ It is worth mentioning the fact SSBP is a rather important effect that arises in a wide range of physical systems modeled by nonlinear equations \cite {Hayata}. \ We would also mention the fact that hysteresis effects associated to bifurcations of stationary solutions are theoretically discussed for BEC's in optical lattice under the effect of a Stark-like external field \cite {Zobay}

The $n$-dimensional linear Schr\"odinger equation with a symmetric double well potential has stationary states of a definite even $\varphi_+$ and odd-parity $\varphi_-$, with associate nondegenerate eigenvalues $\lambda_+ < \lambda_-$. \ However, the introduction of a nonlinear term,  which usually models in quantum mechanics an interacting many-particle system, may give rise to asymmetrical states related to SSBP.

In the semiclassical limit has been proved that the symmetric stable stationary state bifurcates when the adimensional nonlinear parameter $\eta$ takes absolute value equal to the critical value
\begin{eqnarray}
\eta^\star = 2^r /r \, . \label {equation4}
\end{eqnarray}
The parameter $\eta$ is associated with the coupling factor of the nonlinear perturbation by 
\begin{eqnarray}
\eta := \eta (t) = cg_{r+1} (t)/{\omega} \label {equation5}
\end{eqnarray}
and it is the effective nonlinear coupling factor, where $\omega$ is the (half of the) splitting between the two levels 
\begin{eqnarray}
\omega = \frac 12 (\lambda_- - \lambda_+)
\end {eqnarray}
and $c$ is the constant given by 
\begin{eqnarray*}
c = \langle \varphi_R , |\varphi_R |^{2r} \varphi_R \rangle  = \langle \varphi_L , |\varphi_L |^{2r} \varphi_L \rangle \ . 
\end{eqnarray*}
Here, $\varphi_R$ and $\varphi_L$ are the normalized right and left hand-side vectors 
\begin{eqnarray*}
\varphi_R ={(\varphi_+ + \varphi_-)}/{\sqrt {2}} 
\end {eqnarray*}
and 
\begin {eqnarray*} 
\varphi_L =  {(\varphi_+ - \varphi_-)}/{\sqrt {2}},
\end{eqnarray*}
usually named \emph {single-well states} because they are localized on only one well. \ In fact, in the semiclassical limit (or also for large distance between the two wells) the splitting $\omega$ is exponentially small, as $\hbar$ goes to zero, and the supports of the two vectors $\varphi_R$ and $\varphi_L$ don't overlap up to an exponentially small term..

By adopting the two level approximation then the wave function $\psi (x,t)$ is a linear combination of the right and left hand-side vectors
\begin{eqnarray*}
\psi (x,t) = a_R(t) \varphi_R (x) + a_L(t) \varphi_L(x) 
\end {eqnarray*}
where we set
\begin{eqnarray*}
a_R  = p e^{i\alpha }, \ a_L = q e^{i\beta }, \ p^2+q^2 =1 \, . 
\end{eqnarray*}
Defining the relative phase difference $\theta = \alpha - \beta$ and the imbalance function $z=p^2 - q^2$, and rescaling the time as $\tau = \omega t /\hbar$ (hence, the linear beating period takes the value $\pi$), then equation (\ref {equation2}) can be written in the Hamiltonian form
\begin{eqnarray*}
\frac {\partial \theta }{\partial \tau} = \frac {\partial {\mathcal H}}{\partial z} \ \ \mbox { and } \ \ \frac {\partial z }{\partial \tau}  = - \frac {\partial {\mathcal H} }{\partial \theta}
\end{eqnarray*}
with Hamiltonian function
\begin{eqnarray*}
{\mathcal H} = 2 \sqrt {1-z^2} \cos \theta - \eta \frac {(1+z)^{r+1} + (1-z)^{r +1}}{2^r (r +1)} \, . 
\end{eqnarray*}
The energy functional ${\mathcal E}$ associated to the nonlinear Schr\"odinger equation (\ref {equation2}) and written in the two level approximation takes the form ${\mathcal E} = \Omega - \frac 12 \omega {\mathcal H}$, where $\Omega  = \frac 12 (\lambda_- + \lambda_+)$ is the mean value between the two energy levels.

\begin{center}
\begin{figure}
\includegraphics[height=5cm,width=7cm]{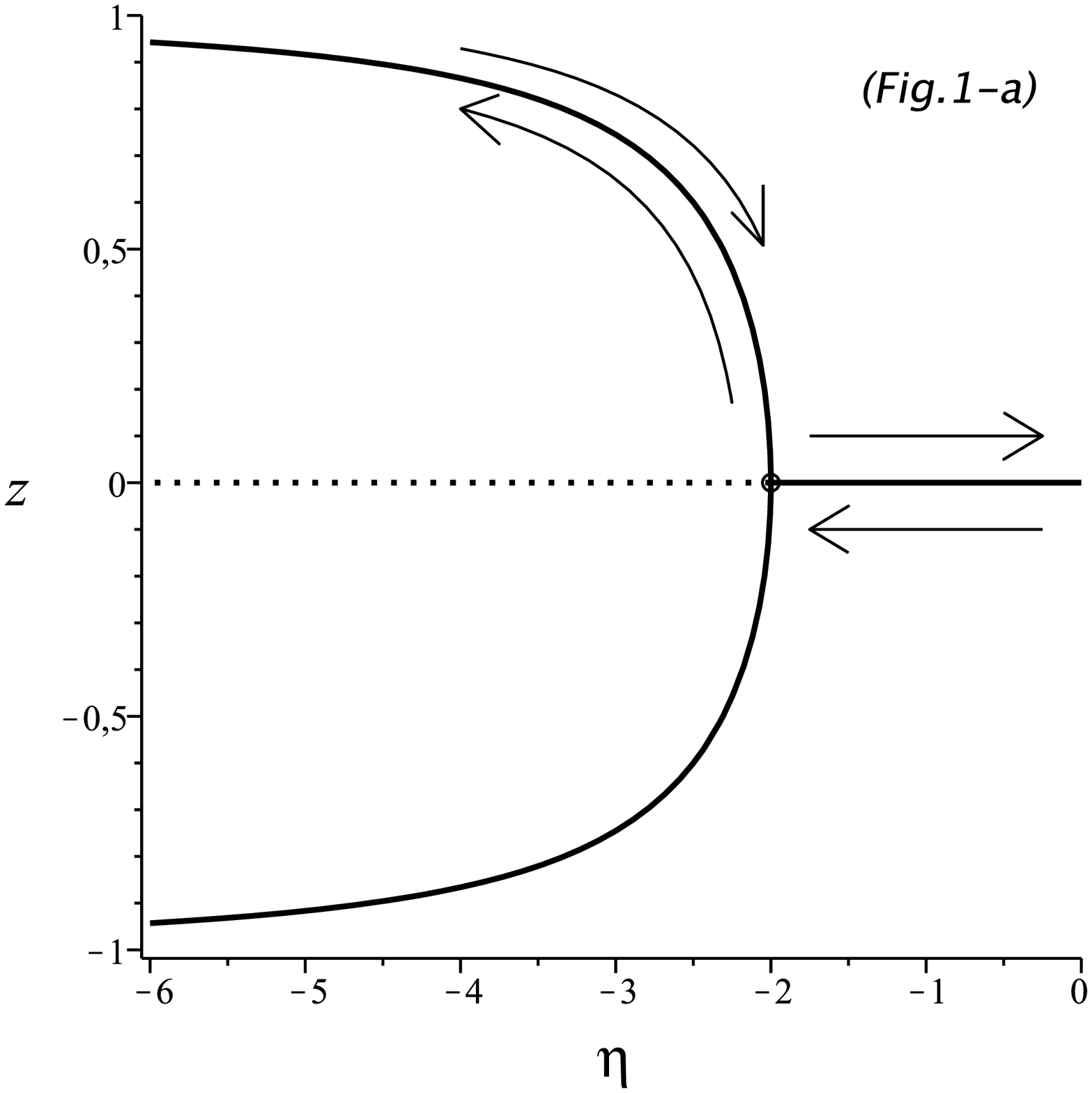}
\includegraphics[height=5cm,width=7cm]{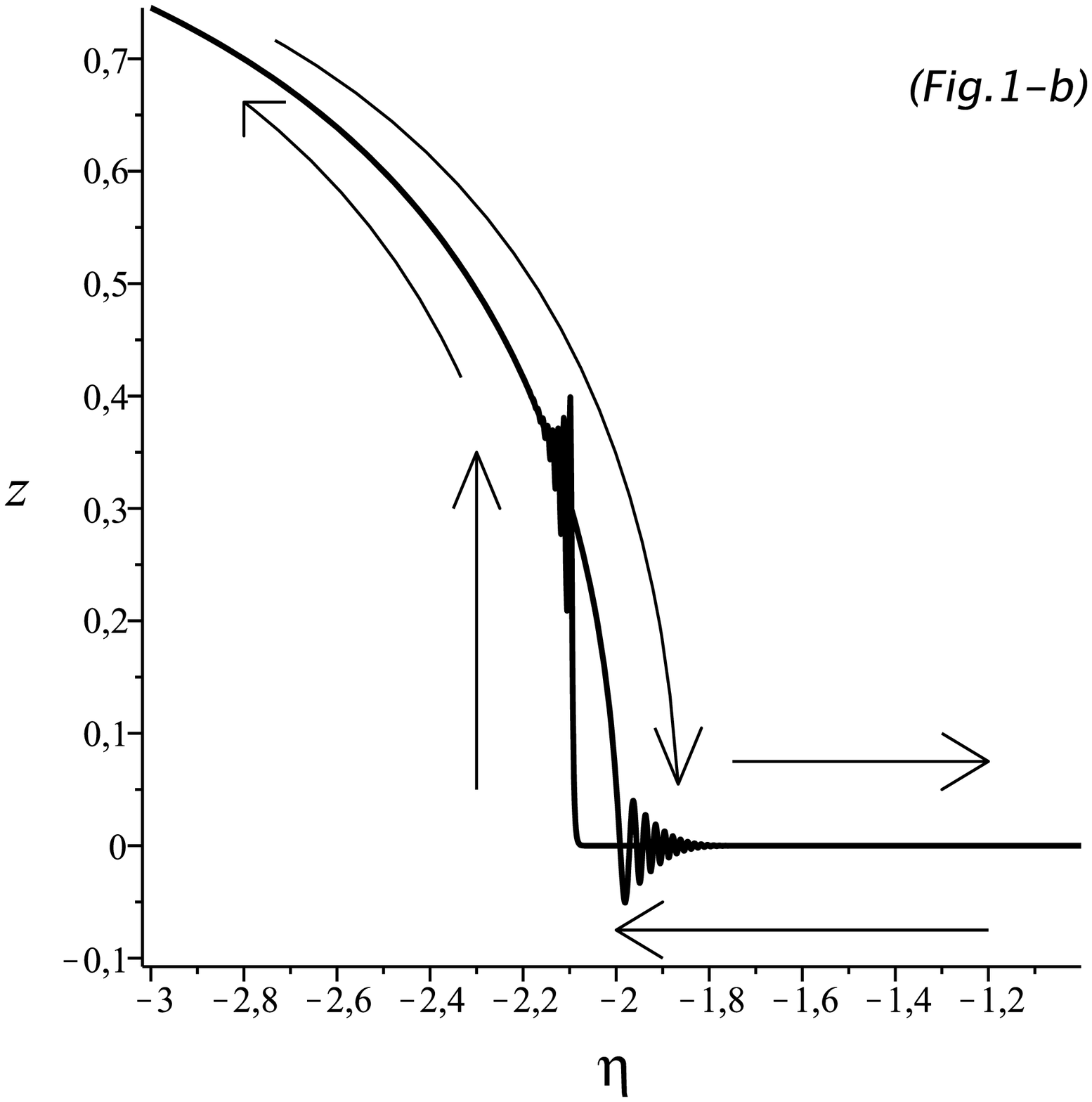}
\caption{\label {Fig1} In this figure we consider the case $r=1 < r_{threshold}$. \ In panel (a) we plot the bifurcation diagram where full lines represents stable stationary solutions and dot lines represent unstable states. \ The arrows represent the "path" of a solution initially close to the symmetric ground state. \ A state initially close to the stable symmetric stationary solution makes experience of a bifurcation effect at $|\eta |= \eta^\star =2$ and it follows the new asymmetric stable stationary solution on one of the branches for $|\eta |  >\eta^\star$. \ When $|\eta |$ moves from values bigger than $\eta^\star$ to values less than $\eta^\star$ then the state returns to be close to the stationary symmetric state without exhibiting hysteresis phenomenon. \ In panel (b) we show the numerical solution of the equations (\ref {equation6}) and (\ref {equation7}) for $r=1$, $\nu =0.5$, $z_0=0.01$ and $\theta_0=0$.}
\end{figure}
\end{center}

We consider at first the case $r\le r_{threshold}$. \ Since $\eta$ takes negative values then the nonlinear ground state is a stable symmetric state for any $|\eta | < \eta^\star$. \ At $|\eta |=\eta^\star$ it bifurcates and we observe also an exchange of the stability properties: for $|\eta |$ larger than $\eta^\star$ then the symmetric stationary state becomes unstable and the new asymmetrical states are stable (see Fig. \ref {Fig1}-a).

\begin{center}
\begin{figure}
\includegraphics[height=5cm,width=7cm]{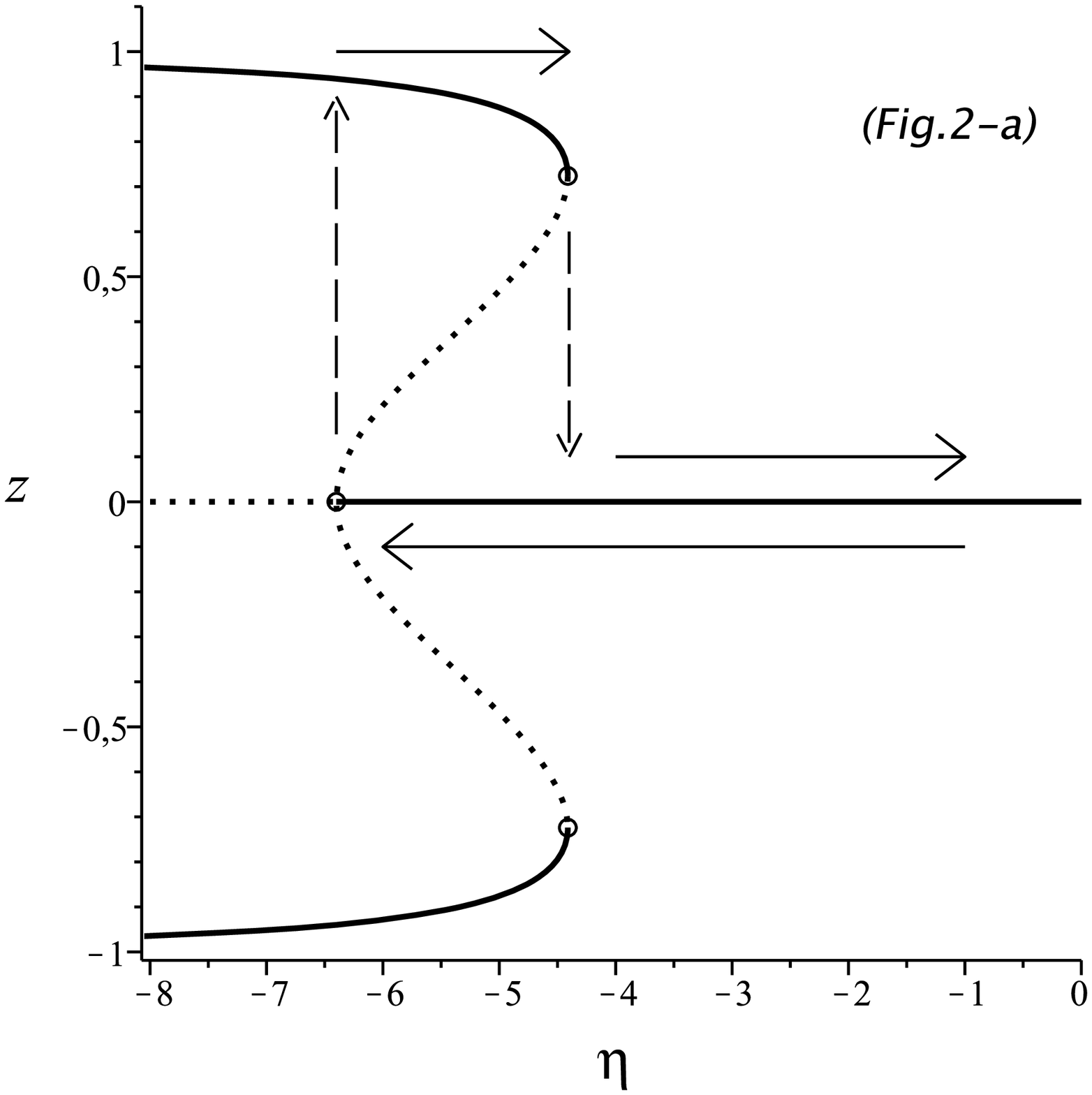}
\includegraphics[height=5cm,width=7cm]{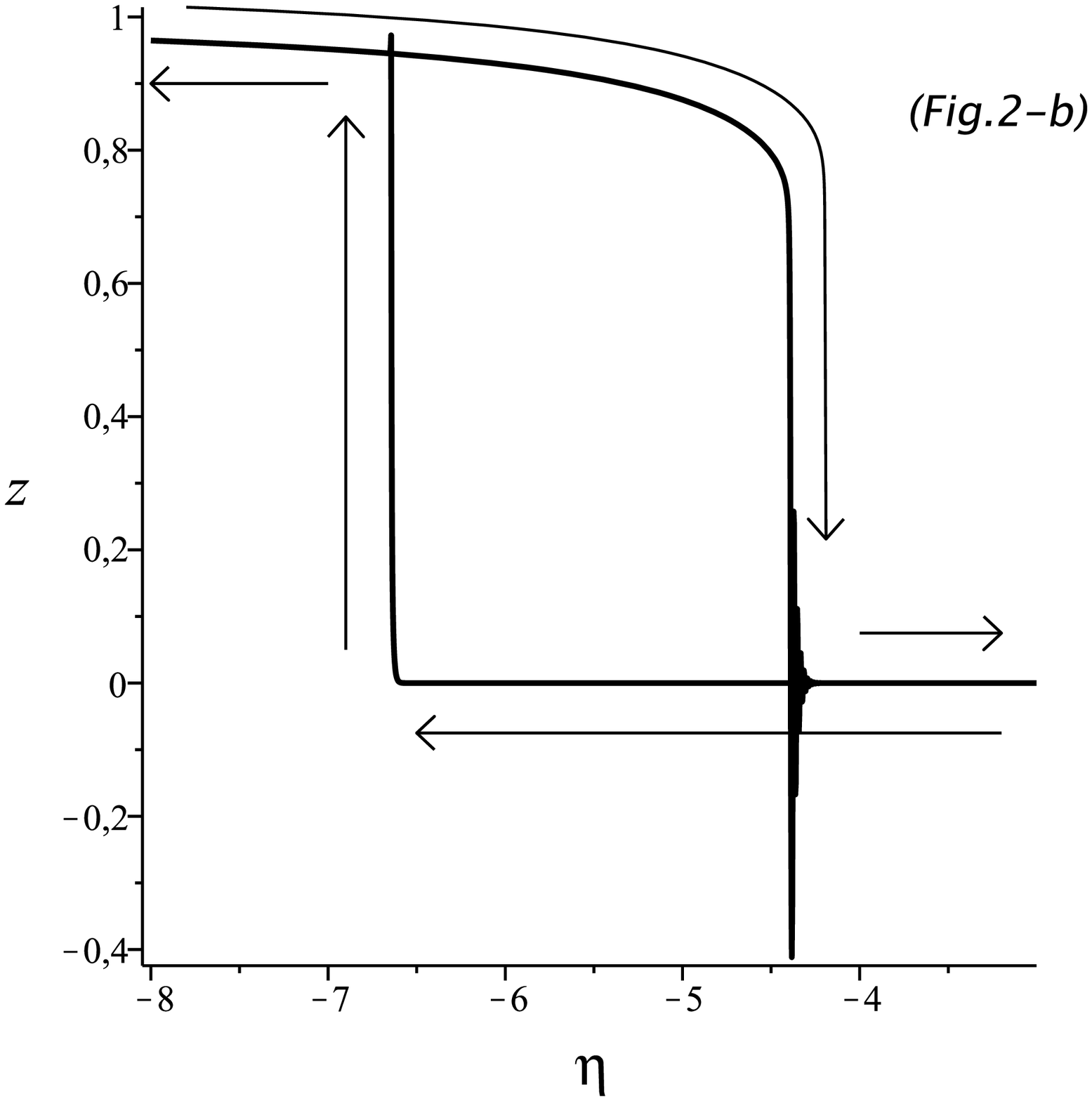}
\caption{\label {Fig2} In this figure we consider the case $r=5 > r_{threshold}$. \ A state initially close to the stable symmetric stationary solution jumps to the new asymmetric stable stationary solution on one of the branches for $|\eta |  >\eta^\star = 6.4$. \ When $|\eta |$ moves from values bigger than $\eta^\star$ to values less than $\eta^\star$ the state does not return to be close to the stationary state, but it exhibits an hysteresis phenomena for $|\eta |$ between $\eta^\star$ and $\eta^+$.  \ Broken arrow are associated to the damping effect which forces the state to collapse on the ground state after some damped oscillations. \ In the right panel we show the numerical solution of the equation (\ref {equation6}) and (\ref {equation7}) for $r=5$, $\nu =0.5$, $z_0=0.01$ and $\theta_0=0$.}
\end{figure}
\end{center}

On the other side, for $r > r_{threshold} $,  then a couple of saddle-node bifurcations, associated to new asymmetrical stationary states, sharply appears when $|\eta |$ is equal to a given value $\eta^+$ such that $\eta^+ < \eta^\star$ \cite {eta}; then, for increasing values of $|\eta |$, the two unstable solutions disappear at $|\eta |=\eta^\star $ showing a subcritical pitch-fork bifurcation (see Fig. \ref {Fig2}-a). 

In such a scenario, that is the sharp appearance of new asymmetrical stationary solutions fully localized on a single well when $r > r_{threshold} $, a new relevant effect occurs: namely we expect to observe hysteresis effect when we adiabatically changes the effective coupling factor $\eta$ such that its absolute value moves from values less than $\eta^\star$ to values bigger than $\eta^\star$ and then it goes back to its initial value. \ To this end we consider a state that, in the $(z,\theta )$-representation, is initially close to the symmetric stationary state: that is its initial condition corresponds to $z_0 \approx 0$ and $\theta_0 \approx 0$. 

In the case $r \le r_{threshold}$ then, as $|\eta|$ increases, the state remains close to the symmetric stationary state for any $|\eta | < \eta^\star$, at $|\eta | = \eta^\star$ it makes experience of a bifurcation and it follows one of two branches for $|\eta |> \eta^\star$. \ When $|\eta|$ decreases from values bigger than $\eta^\star$ to values less than $\eta^\star$ then such a path is reversed and the state returns close to the initial symmetric stationary state when $|\eta|$ returns to its initial value such that $|\eta |< \eta^\star$ (see the path indicated by the arrows in Fig. \ref {Fig1}-a). 

On the other side, if $r>r_{threshold}$ then the state still remains close to the initial stable stationary state $(z_0, \theta_0 )$ for any $|\eta |<\eta^\star$, but at $|\eta |=\eta^\star$ we don't have a smooth bifurcation and for $|\eta | > \eta^\star$ the state starts to oscillate around the stable asymmetric stationary solution localized on only one of the two wells. \ As $|\eta |$ decreases from values bigger than $\eta^\star$ to values between $\eta^+$ and $\eta^\star$ the previous path is not reversed. \ In fact, the state continues to oscillate around the stable asymmetric stationary solution until $|\eta|$ reaches the value $\eta^+$. \ Then, while $\eta$ is returning to its initial value it takes the value $|\eta |=\eta^+$, for which the asymmetrical stable stationary states disappear,  and the wavefunction starts to exhibit a wide oscillating motion around the symmetrical stationary solution corresponding to $z=0$. \ If we introduce a small damping effect then such  oscillating motions are damped and the state will stay close to the symmetric stationary solution. \ In conclusion we can see in such a scenario that an hysteresis effect appears for values of $|\eta |$ between $\eta^+$ and $\eta^\star$ (see the path indicated by arrows in Fig. \ref {Fig2}-a).

In fact, such an hysteresis effect becomes more evident by adding a damping term which forces the state to collapse to the ground state. \ Actually, in physical systems we should expect to take into account a certain amount of damping due to the incoherent exchange of normal atoms. \ In particular, an accepted model for damped two-components BECs has been introduced by \cite {Marino} and it reads as
\begin{eqnarray}
\frac {\partial z}{\partial \tau} = - \frac {\partial {\mathcal H}}{\partial \theta} - \nu \frac {\partial \theta}{\partial \tau} := - \sqrt {1-z^2} \sin \theta - \nu \frac {\partial \theta}{\partial \tau} \label {equation6}
\end {eqnarray}
where $\nu >0$ is the damping constant, and
\begin{eqnarray}
\frac {\partial \theta}{\partial \tau} = \frac {\partial {\mathcal H}}{\partial z}  := -\frac {2z\cos \theta}{\sqrt {1-z^2}} - \frac {\eta}{2^r} \left [ (1+z)^r - (1-z)^r \right ]  \label {equation7}
\end {eqnarray}
In Fig. \ref {Fig1}-b (for $r=1$) and Fig. \ref {Fig2}-b (for $r=5$) we plot the numerical solutions of this dynamical system, where we assume the initial condition $\theta_0 = 0$ and $z_0 =0.01$ closed to the symmetric stationary solution, and for times $\tau \in [0,T]$ where $T=4000$. \ The damping factor is chosen to be $\nu =0.5$ and the time dependent function $\eta$ has the following form
\begin {eqnarray*}
\eta (\tau ) = -1-2 \left [ 1- |2\tau /T - 1 |\right ] ,\ \ \mbox { if } \ r=1 ,
\end {eqnarray*}
and
\begin {eqnarray*}
\eta (\tau ) = -3-5 \left [ 1- |2\tau /T - 1 |\right ] ,\ \ \mbox { if } \ r=5.
\end {eqnarray*}

As predicted by means of the previous analysis on the bifurcation of the stationary solutions, it appears that for $r=1$ no hysteresis effect occurs; while, for $r=5$ the hysteresis effect occurs for $|\eta|$ between $\eta^+$ and $\eta^\star$. \ Oscillations of the state that occur when $|\eta |$ becomes less than $\eta^\star$ are damped because of the damping factor. \ It is worth mentioning also the fact that the delay observed in the case $r=1$, when the absolute value of $\eta$ becomes larger than the branch point $\eta^\star$, is not a consequence of some hidden physical effects but it comes from the singularity associated to the branch point. \ In fact, for larger values of $T$ then $\eta $ is almost constant around the branch point and this delay disappear.

In summary, we have shown that in a damped BEC with a double-well trapping potential the nonlinear term coming from a $(r+1)$-body interaction, for $r$ bigger than $4$, may give rise to an hysteresis effect when the corresponding coupling factor adiabatically changes. \ The modulation of the coupling factor can be performed by tuning the condensate population by means of suitable external fields. \ Such an hysteresis effect has not theoretically predicted by means of the well-known Gross-Pitaevskii equation, in which only binary contact potentials are considered.

\end {document}